\documentclass[12pt]{article}

\newcommand{\be}{\begin{equation}}
\newcommand{\ee}{\end{equation}}
\newcommand{\bea}{\begin{eqnarray}}
\newcommand{\eea}{\end{eqnarray}}

\begin{document}
\title{Implications of $\eta'$ Coupling In The Chiral Constituent Quark Model}
\author{Harleen Dahiya$^a$, Manmohan Gupta$^a$ and J.M.S. Rana$^{b}$     \\
{\small {\it $^a$Department of Physics, Centre of Advanced Study
in Physics,}} \\ {\small{\it Panjab University, Chandigarh-160
014, India.}} \\ {\small{\it $^b$ Department of Physics, H.N.B.
Garhwal University, SRT Campus }}\\
 {\small{\it Badshahithaul(Tehri-Garhwal), India.}}}
\maketitle

\begin{abstract}

Using the latest data pertaining to $\bar u-\bar d$ asymmetry and
the spin polarization functions, detailed implications of the
possible values of the coupling strength of the singlet Goldstone
boson $\eta'$ have been investigated in the $\chi$CQM with
configuration mixing. Using $\Delta u$, $\Delta_3$, $\bar u-\bar
d$ and $\bar u/\bar d$, the possible ranges of the coupling
parameters $a$, $\alpha^2 a$, $\beta^2 a$ and $\zeta^2 a$,
representing respectively the probabilities of fluctuations to
pions, $K$, $\eta$ and $\eta^{'}$, are found. To further constrain
the coupling strength of $\eta'$, detailed fits have been obtained
for spin polarization functions, quark distribution functions and
baryon octet magnetic moments. The fits clearly establish that a
small non-zero value of the coupling of $\eta'$ is preferred over
the higher values of $\eta'$ as well as when $\zeta=0$, the latter
implying the absence of $\eta'$ from the dynamics of $\chi$CQM.
Our best fit achieves an overall excellent fit to the data, in
particular for $\Delta u$, $\Delta d$, $\Delta_8$ as well as the
magnetic moments $\mu_{n}$, $\mu_{\Sigma^-}$, $\mu_{\Sigma^+}$ and
$\mu_{\Xi^-}$. The implications of $\eta'$ on the gluon
polarization have also been investigated.

\end{abstract}

\maketitle

The chiral constituent quark model ($\chi$CQM), as formulated by
Manohar and Georgi \cite{{manohar}}, has recently got good deal of
attention \cite{{eichten},{cheng},{song},{johan}} as it is
successful in not only explaining the ``proton spin crisis''
\cite{emc} but is also able to account for the $\bar u-\bar d$
asymmetry {\cite{{nmc},{e866},{GSR}}}, existence of significant
strange quark content $\bar s$ in the nucleon, baryon magnetic
moments {\cite{{eichten},{cheng}}} and hyperon $\beta-$decay
parameters etc..  Further, $\chi$CQM with configuration mixing
(henceforth to be referred as $\chi$CQM$_{{\rm config}}$) when
coupled with the quark sea polarization and orbital angular
 momentum (Cheng-Li mechanism \cite{{cheng}}) as well as
``confinement effects'' is able to give an excellent fit \cite{hd}
to the baryon magnetic moments and a perfect fit for the violation
of Coleman Glashow sum rule.

The key to understand the ``proton spin problem'', in the
$\chi$CQM formalism \cite{cheng}, is the fluctuation process $
q^{\pm} \rightarrow {\rm GB}
  + q^{' \mp} \rightarrow  (q \bar q^{'})
  +q^{'\mp}$, where GB represents the Goldstone boson and $q \bar q^{'}  +q^{'}$
 constitute the ``quark sea'' \cite{cheng,song,johan,hd}.
The effective Lagrangian describing interaction between quarks and
a nonet of GBs, consisting of octet and a singlet, can be
expressed as $ {\cal L}= g_8 {\bf \bar q}\Phi {\bf q} + g_1{\bf
\bar q}\frac{\eta'}{\sqrt 3}{\bf q}= g_8 {\bf \bar
q}\left(\Phi+\zeta\frac{\eta'}{\sqrt 3}I \right) {\bf q}=g_8 {\bf
\bar q}\left(\Phi'\right) {\bf q}$, where $\zeta=g_1/g_8$, $g_1$
and $g_8$ are the coupling constants for the singlet and octet
GBs, respectively, $I$ is the $3\times 3$ identity matrix. The GB
field $\Phi'$ includes the octet and the singlet GBs. The
parameter $a(=|g_8|^2$) denotes the probability of chiral
fluctuation  $u(d) \rightarrow d(u) + \pi^{+(-)}$, whereas
$\alpha^2 a$, $\beta^2 a$ and $\zeta^2 a$ respectively denote the
probabilities of fluctuations $u(d) \rightarrow s + K^{-(0)}$,
$u(d,s) \rightarrow u(d,s) + \eta$,
 and $u(d,s) \rightarrow u(d,s) + \eta^{'}$.

The chiral structure of QCD is known to have intimate connection
with the $\eta$ and $\eta'$ dynamics \cite{etaprob}. Recently, in
the context of $\chi$CQM, Steven D. Bass \cite{bass} has
reiterated in detail the deep relationship of the non-perturbative
aspects of QCD, including gluon anomaly, and the comparatively
large masses of the $\eta$ and $\eta'$ mesons.  Similarly, it has
been shown earlier by Ohta {\it et al.} \cite{ohta} and recently
advocated by Cheng and Li \cite{cheng} that $\eta'$ could play an
important role in the formulation of the $\chi$CQM.  On the other
hand, it has recently been observed on phenomenological grounds
\cite{johan} that the new measurement of both the $\bar u/\bar d$
asymmetry as well as $\bar u-\bar d$ asymmetry by the NuSea
Collaboration \cite{e866} may not require substantial contribution
of $\eta'$. In this context, it therefore becomes interesting to
understand the extent to which the contribution of $\eta'$ is
required in the $\chi$CQM thereby giving vital clues to the
dynamics of non-perturbative regime of QCD.

To study the variation of the $\chi$CQM parameters and the role of
the coupling strength of $\eta'$ in obtaining the fit, one needs
to formulate the experimentally measurable quantities having
implications for these parameters.  The spin structure of a
nucleon is defined as \cite{{cheng},{song},{johan},hd} $\hat B
\equiv \langle B|N|B\rangle,$ where $|B\rangle$ is the nucleon
wavefunction and $N$ is the number operator giving the number of
$q^{\pm}$ quarks.  The contribution to the proton spin in
$\chi$CQM$_{{\rm config}}$ is given by the spin polarizations
defined as $\Delta q=q^+-q^-$. After formulating the spin
polarizations of various  quarks, we consider several measured
quantities which are expressed in terms of the above mentioned
spin  polarization functions. The flavor non-singlet components
$\Delta_3= \Delta u-\Delta d$ and $\Delta_8= \Delta u+\Delta d-2
\Delta s$, usually calculated in the $\chi$CQM, are obtained from
the neutron $\beta-$decay and the weak decays of hyperons. The
flavor non-singlet component $\Delta_3$ is related to the well
known Bjorken sum rule \cite{bjor}. Another quantity which is
usually evaluated is the flavor singlet component $\Delta \Sigma=
\frac{1}{2}(\Delta u+\Delta d+\Delta s)$, in the $\Delta s=0$
limit, this reduces to the Ellis-Jaffe sum rule \cite{ellis}.
Apart from the above mentioned spin polarization we have also
considered the quark distribution functions which have
implications for $\zeta$ as well as for other $\chi$CQM
parameters. For example, the antiquark flavor contents of the
``quark sea'', the deviation of Gottfried sum rule \cite{GSR},
related to the $\bar u(x)$ and $\bar d(x)$ quark distributions,
$\bar u/\bar d$  and the fractions of the quark content defined as
$f_q=\frac{q+\bar q}{[\sum_{q} (q+\bar q)]}$.

With a view to  phenomenologically estimating the coupling
strength of the singlet Goldstone boson $\eta'$ in
$\chi$CQM$_{{\rm config}}$,  we have carried out a detailed
analysis  using the latest data regarding $\bar u-\bar d$
asymmetry, the spin polarization functions and the baryon octet
magnetic moments.  As a first step of the analysis, we have found
from broad considerations the required ranges of these parameters
using the data pertaining to $\Delta u$, $\Delta_3$, $\bar u-\bar
d$, $\bar u/\bar d$ etc.. After obtaining the ranges, analysis has
been carried out corresponding to four different sets of the
$\chi$CQM parameters within the ranges. In the first case, the
pion fluctuation parameter $a$ is taken as 0.1, whereas $\Delta
u$, $\Delta_3$, $\bar u-\bar d$, $\bar u/\bar d$ etc. are fitted
by treating the other three parameters to be free. This analysis
yields $|\zeta|=0.65$, $\alpha=0.4$ and $\beta=0.7$ and is
referred to as Case I. A similar analysis has also been also been
carried out by taking $a=0.1$, $|\zeta|=0.70$, $\alpha=0.4$ and
$\beta=0.6$ and is referred to as Case II. Our best fit (Case IV)
is obtained by varying $a$, $\zeta$ and $\alpha$, the parameter
$\beta$ is taken to be equal to $\alpha$ and the best fit values
of the parameters are $a=0.13$, $|\zeta|=0.10$,
$\alpha=\beta=0.45$. We have also carried out a fit where there is
no contribution of the singlet GB ($\zeta=0$) and $a$, $\alpha$ as
well as $\beta$ are treated free, yielding $a=0.14$, $\alpha=0.4$
and $\beta=0.2$ and referred to as Case III.

In Table I, we have presented the results of our fits mentioned
above. A comparison of all the fits clearly shows that our best
fit is not only better than other fits carried out here but also
provides an excellent overall fit to the data particularly in the
case of $\Delta u$, $\Delta d$, $\Delta_8$, $\mu_{n}$,
$\mu_{\Sigma^-}$, $\mu_{\Sigma^+}$, $\mu_{\Xi^o}$ and
$\mu_{\Xi^-}$. It needs to be mentioned that $\Delta_8$ cannot be
fitted for  "higher values" of $|\zeta|$ even after scanning the
entire parameter space for $a$, $\alpha$ and $\beta$, suggesting
that only the lower values of $|\zeta|$ are compatible with data.
In Table II, we have presented the results corresponding to quark
distribution functions. In this case also the fit for the lower
values of $|\zeta|$ is better as compared to the higher values. It
is interesting to observe that even for a small deviation in the
value of $\zeta$, $f_s$ gets affected significantly, therefore a
measurement of $f_s$ would give a very strong signal about the
coupling strength of $\eta'$ in the $\chi$CQM. It may be mentioned
that our conclusion regarding the small but non-zero value of
$|\zeta|$ being preferred over $\zeta=0$  is not only in agreement
with latest $\bar u-\bar d$ measurement \cite{e866} but is also in
agreement with the conclusions of Ohlsson {\it et. al.}
\cite{johan}.

To further understand the coupling strength of $\eta'$ in
$\chi$CQM, we have carried out in Case III an analysis where the
contribution of $\eta'$ is taken to be zero. Interestingly, we
find that the fit obtained in this case cannot match our best fit
even if the other parameters are treated completely free,
suggesting that the $\zeta=0$ case can be is excluded
phenomenologically. This conclusion regarding the exclusion of the
singlet GB also looks to be in agreement with the theoretical
considerations based on the arguments of Cheng and Li \cite{cheng}
and those of S. Bass \cite{bass}. It seems that the
phenomenological analyses of spin polarization functions, quark
distribution functions and baryon octet magnetic moments, strongly
suggest a small but non-zero value of $|\zeta|$ within the
dynamics of chiral constituent quark model, suggesting an
important role for $\eta'$ in the non-perturbative regime of QCD.

\vskip .2cm
 {\bf ACKNOWLEDGMENTS}\\
H.D. would like to thank DST, Government of India and the
organizers of SPIN2006, Kyoto University, for financial support.

\begin{table}
\begin{tabular}{lrrrrr} \hline
Parameter & Data  & \multicolumn{4}{c}{$\chi$CQM$_{{\rm config}}$}
\\  \cline{3-6}
& & Case I& Case II & Case III & Case IV \\ \cline{3-6} & &
$|\zeta|=0.65$ & $|\zeta|=0.70$ & $\zeta=0$ & $|\zeta|=0.10$\\
 &     & $a=0.1$ &
$a=0.1$ & $a=0.14$ & $a=0.13$\\ & & $\alpha=0.4$ & $\alpha=0.4$ &
$\alpha=0.4$ & $\alpha=0.45$ \\ & & $\beta=0.7$ & $\beta=0.6$ &
$\beta=0.2$ & $\beta=0.45$
\\
 \hline

$\Delta u$ & 0.85 $\pm$ 0.05 \cite{emc} & 0.947 & 0.955 & 0.925 &
0.913
\\ $\Delta d$ & $-$0.41  $\pm$ 0.05 \cite{emc}   & $-$0.318 &
$-0.312$& $-$0.352& $-$0.364 \\ $\Delta s$ &$-$0.07  $\pm$ 0.05
\cite{emc} &$-$0.02&$-$0.02 &$-$0.02 &$-$0.02 \\ $\Delta_3$ &
1.267 $\pm$ 0.0035 \cite{PDG} & 1.267 & 1.267 & 1.267 & 1.267\\
$\Delta_8$ & 0.58  $\pm$ 0.025 {\cite{PDG}} &0.67 & 0.68& 0.61 &
0.59 \\ $\Delta \Sigma$ & 0.19 $\pm$ 0.025 {\cite{PDG}}& 0.31&
0.31 &0.28 & 0.27
\\ \hline $\mu_{p}$ & 2.79$\pm$0.00 {\cite{PDG}} & 2.80 & 2.80 & 2.83 & 2.81 \\
 $\mu_{n}$ & $-1.91\pm$0.00 {\cite{PDG}}&  $-$1.99 & $-$2.00 &
$-$2.16 & $-$1.96  \\ $\mu_{\Sigma^-}$ & $-1.16\pm$0.025
{\cite{PDG}}& $-$1.20 & $-$1.21& $-$1.32& $-$1.19 \\
$\mu_{\Sigma^+}$ & 2.45$\pm$0.01 {\cite{PDG}} & 2.43& 2.42& 2.53 &
2.46
\\ $\mu_{\Xi^o}$ & $-1.25\pm$0.014 {\cite{PDG}} &  $-$1.24 & $-$1.23 & $-$1.33 & $-$1.26 \\
$\mu_{\Xi^-}$ & $-0.65\pm$0.002 \cite{PDG} & $-$0.56 &$-$0.57&
$-$0.67 & $-$0.64 \\ \hline
\end{tabular}

 \caption{The calculated values of the spin polarization
functions and  baryon octet  magnetic moments for different cases.
The value of the mixing angle $\phi$ is taken to be $20^o$. }
\label{spin}
\end{table}

\begin{table}
\begin{tabular}{lcrrrr} \hline

Parameter & Data  & \multicolumn{4}{c}{$\chi$CQM} \\ \cline{3-6} &
& Case I& Case II & Case III & Case IV \\ \cline{3-6} & &
$|\zeta|=0.65$ & $|\zeta|=0.70$ & $\zeta=0$ & $|\zeta|=0.10$\\ & &
$a=0.1$ & $a=0.1$ & $a=0.14$& $a=0.13$ \\ & & $\alpha=0.4$ &
$\alpha=0.4$ & $\alpha=0.4$ & $\alpha=0.45$ \\ & & $\beta=0.7$ &
$\beta=0.6$ & $\beta=0.2$ & $\beta=0.45$
\\ \hline $\bar u$ & $-$ & 0.168 & 0.167&0.250 & 0.233 \\

$\bar d$ & $-$ &  0.288 & 0.293&  0.366& 0.350 \\

$\bar s$ & $-$   &   0.108 & 0.104& 0.07& $0.07$ \\

$\bar u-\bar d$ & $-0.118 \pm$ 0.015 \cite{e866} & $-0.120$&
$-0.127$ & $-0.116$ & $-0.117$ \\

$\bar u/\bar d$ & 0.67 $\pm$ 0.06 {\cite{e866}}  & 0.58 &0.57
&0.68& 0.67  \\

$I_G$ & 0.254  $\pm$ 0.005  & 0.253 &0.248&0.255 & 0.255  \\

$f_u$ &$-$   &   0.655 & 0.654 &0.677& 0.675 \\

$f_d$ &$-$ &  0.442 &0.445& 0.470 & 0.466 \\

$f_s$ &  0.10 $\pm$ 0.06 {\cite{ao}}  &  0.061 & 0.058 &0.039 &
0.039
\\

$f_3$  &$-$ & 0.213 & 0.209 & 0.207 & 0.209\\

$f_8$  &$-$ & 0.975 &0.982&  1.07& 1.06 \\

$f_3/f_8$ & 0.21 $\pm$ 0.05 {\cite{cheng}}  &  0.22 & 0.21 & 0.19
& 0.20 \\ \hline

\end{tabular}

\caption{The calculated values of the quark flavor distribution
functions for different cases.} \label{quark}
\end{table}

\end{document}